\documentclass[twocolumn,prl,showpacs,aps]{revtex4-1}

\usepackage[english]{babel}
\usepackage[ansinew]{inputenc}
\usepackage{times}
\usepackage{graphicx}
\usepackage{graphics}
\usepackage{amsmath}
\usepackage{amsfonts}
\usepackage{amssymb}
\usepackage{dsfont}
\usepackage{epstopdf}
\usepackage{makeidx}
\usepackage{subfigure}
\usepackage[usenames, dvipsnames]{color}
\usepackage{pgf}
\usepackage{bm}
\usepackage{tikz} 
\usepackage[normalem]{ulem}

\newcommand{\be}{\begin{equation}}
\newcommand{\ee}{\end{equation}}
\newcommand{\bea}{\begin{eqnarray}}
\newcommand{\eea}{\end{eqnarray}}

\makeindex
\begin{document}
\title{Spontaneous natural optical activity in disordered media}
\author{F. A. Pinheiro$^{1}$, V. A. Fedotov$^{2}$, N. Papasimakis$^{2}$, N. I. Zheludev$^{2,3}$}
\address{$^{1}$ Instituto de F\'{i}sica, Universidade Federal do Rio de Janeiro, Rio de Janeiro-RJ, 21941-972, Brazil}
\address{$^{2}$ Optoelectronics Research Centre and Centre for Photonic Metamaterials, University of Southampton, United Kingdom}
\address{$^{3}$ Centre for Disruptive Photonic Technologies, School of Physical and Mathematical Sciences and The Photonics Institute, Nanyang Technological University, Singapore}

\begin{abstract}
We demonstrate natural optical activity in disordered ensembles of non-chiral plasmonic resonators. We show that the statistical distributions of rotatory power and spatial dichroism are strongly dependent on the scattering mean free path in diffusive random media. This result is explained in terms of the intrinsic geometric chirality of disordered media, as they lack mirror symmetry. We argue that chirality and natural optical activity of disordered systems can be quantified by the standard deviation of both rotatory power and spatial dichroism. Our results are based on microscopic electromagnetic wave transport theory coupled to vectorial Green's matrix method for pointlike scatterers, and are independently confirmed by full-wave simulations.

\end{abstract}
\maketitle
%
The concept of chirality, introduced by Lord Kelvin to designate any geometrical object or ensemble of points that lack mirror symmetry, pervades the natural world. The DNA double-helix structure and the related long-standing enigma of the origin of homochirality of life are just some examples of this fundamental concept in science~\cite{wagniere}.
Since the pioneer work by Pasteur on the resolution of tartaric acid, the development of efficient synthesis processes of naturally occurring chiral molecules is of utmost importance in chemistry, biology, and medicine since chiral systems have widespread applications in pharmaceuticals, asymmetric synthesis, and catalysis~\cite{wagniere}. In addition to natural chiral systems, designing and characterizing artificial chiral structures constitute an intense research activity in recent years~\cite{guerrero}. Chiral metamaterials are striking examples of artificial media without mirror symmetry~\cite{rogacheva}; they have been applied to generate novel optical properties such as negative index media~\cite{pendry,plum2009}, broadband circular polarizers~\cite{gansel}, and enantioselective photochemistry~\cite{tang2011}. Another possibility of artificially generating chiral systems is to assembly achiral plasmonic nanoparticles in chiral geometries, such as helices, pyramids, and twisted nanorod pairs~\cite{zheludev2003,zheludev2006,valev2013,kuzuk,shen2013,hentschel,zan,ferry}. Remarkably, natural optical activity in these systems can be considerably larger than in molecular ones. On a more fundamental side, the conditions for observing natural optical activity, involving the preservation of the helicity of light, have been recently revisited~\cite{corbaton1,corbaton2,corbaton3}.

Despite all efforts to understand the optical properties of naturally occurring chiral media and to design artificial ones, there is an important yet overlooked class of chiral systems: disordered structures. Such systems have neither center nor plane of symmetry, so that they should exhibit natural optical activity as any other chiral medium. Interestingly, there is some previous experimental evidence of natural optical activity in random media, but this effect has never been attributed to the presence of disorder~\cite{hadley,silverman,drachev}; instead, alternative explanations, such as surface contamination by unwanted chiral substances, have been put forward~\cite{hadley,silverman}.

\begin{figure}[h!]
	\centering
	\includegraphics[width=0.4\textwidth]{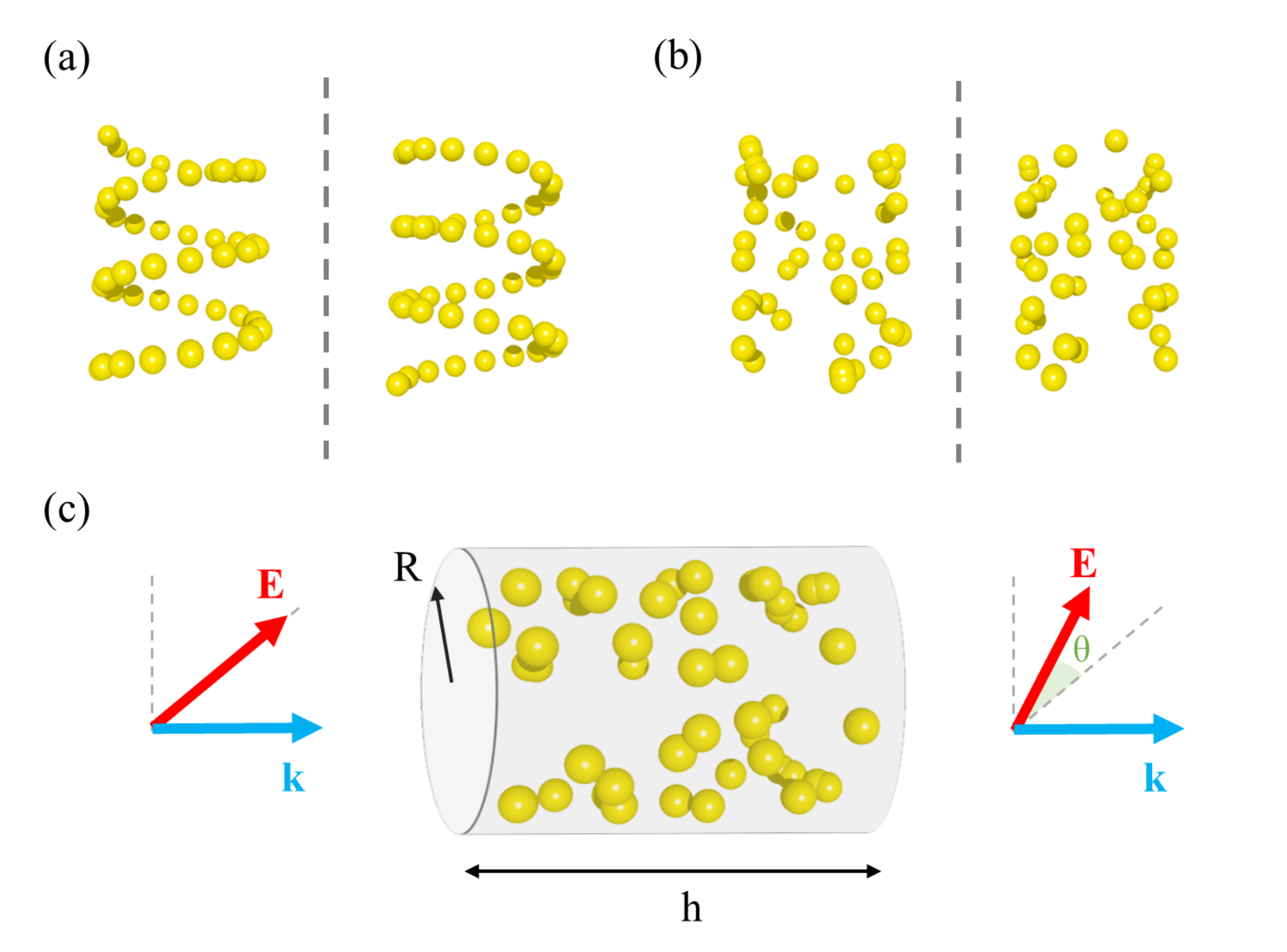}
	\caption{Examples of chiral ensembles: dipole scatterers arranged in a helix (a) and in a random configuration (b); (c) Schematic representation of polarisation rotation in a disordered medium. } \label{fig1}
\end{figure}

Here, we demonstrate natural optical activity due to intrinsic geometric chirality in disordered, diffusive scattering systems. By means of microscopic electromagnetic wave transport theory, we derive an expression for the rotatory power and spatial dichroism of a medium consisting of randomly distributed pointlike scatterers (see Fig. 1). We perform a statistical analysis of natural optical activity in random media to conclude that the standard deviation of both rotatory power and spatial dichroism are strongly dependent on the density of scatterers and the scattering mean free path. Finally, we show that natural optical activity in disordered ensembles can be more that an order of magnitude higher than in the archetypical helical configurations \cite{zan}.

In the following we describe the derivation of the expression for natural optical activity in disordered media using the well established microscopic theory of vector wave diffusion~\cite{pingboek}, which
finds its equivalent in electronic transport theory~\cite{mahan}.
The second-rank tensor ${\mathbf d}_{{\mathbf p}}$ represents the specific intensity
of the diffuse radiation
in the direction ${\mathbf p}$.
It can be expressed as~\cite{mahan}:
\begin{equation}
{\mathbf d}_{{\mathbf p}} ({\mathbf q})
= i \left[ {\mathbf G}_{{\mathbf p}} - {\mathbf G}^{*}_{{\mathbf p}} \right]
-i{\mathbf G}_{{\mathbf p}}  \cdot {\mathbf \Gamma}_{{\mathbf p}} ({\mathbf q})
\cdot
{\mathbf G}^{*}_{{\mathbf p}},
\label{tensord}
\end{equation}
with ${\mathbf G}_{\mathbf p} =
[\omega^2 - p^2 +{\mathbf p} {\mathbf p} - {\mathbf \Sigma}_{{\mathbf p}}]^{-1}$
the Dyson Green's tensor.
For a low density $n$ of the scatterers
 the (second rank)
mass-operator ${\mathbf \Sigma}_{{\mathbf p}}$
is related to the T-matrix of one
independent scatterer~\cite{pingboek},
\begin{equation}
{\mathbf \Sigma}_{{\mathbf p}} = n {\mathbf T}_{{\mathbf p} {\mathbf p}}
 \label{boltz1},
\end{equation}
where we have implicitly
applied rotational averaging of Eq. (\ref{boltz1}). In Eq. (\ref{tensord}) the second-rank tensor ${\mathbf \Gamma}_{{\mathbf p}}$
describes the angular dependence of the diffuse energy flow. It obeys the integral equation~\cite{mahan}
\begin{equation}
{\mathbf \Gamma}_{{\mathbf p}} ({\mathbf q})
= {\mathbf L}_{\mathbf p} ({\mathbf q})
+ \sum_{{\mathbf p}^{\prime}} {\mathbf \Gamma}_{{\mathbf p}^{\prime}}({\mathbf q})
\cdot {\mathbf G}_{{\mathbf p}'} \otimes {\mathbf G}^{*}_{{\mathbf p}'}
\cdot {\mathbf U}_{{\mathbf p}{\mathbf p}'},
\label{gama1}
\end{equation}
where the current tensor is
$ {\mathbf L}_{{\mathbf p}}({\mathbf q}) \equiv 2 ({\mathbf p} \cdot {\mathbf q})\, {\mathbf U}
- {\mathbf p} {\mathbf q} - {\mathbf q} {\mathbf p}$ (${\mathbf U}$ being the unit matrix), and
the (four-rank) irreducible vertex is related, after rotational averaging, to the T-matrix of one independent scatterer according to~\cite{pingboek}
\begin{equation}
{\mathrm U}_{{\mathbf p} {\mathbf p}'}({\mathbf q}) =
n {\mathbf T}_{{\mathbf p} + {\mathbf p}'+} \otimes {\mathbf T}^*_{{\mathbf p}-{\mathbf p}'-}\ ,   \label{vertexu}
\end{equation}
with ${\mathbf p} \pm \equiv {\mathbf p} \pm \frac12 {\mathbf q}$.
For isotropic, achiral media subject to weak disorder,
the well known Boltzmann results applies,
$ {\mathbf \Gamma}_{ij {\mathbf p}} ({\mathbf q}) = ( {\mathbf p} \cdot {\mathbf q})
 {\mathbf \gamma}_{0} \delta_{ij}$ with ${\mathbf \gamma}_{0} = 2/(1 - \langle \cos \theta \rangle)$.

In isotropic chiral media, the tensor ${\mathbf \Gamma}_{{\mathbf p}}$ (neglecting longitudinal terms)
takes the form~\cite{josa}:
\begin{equation}
{\mathbf \Gamma}_{{\mathbf p}} ({\mathbf q})
= {\mathbf \gamma}_{0} ({\mathbf p} \cdot {\mathbf q}) {\mathbf \Delta}_{{\mathbf p}}
+ {\mathbf \gamma}_{C} ({\mathbf p} \cdot {\mathbf q}) {\mathbf \Phi}_{{\mathbf {\widehat p}}},
\label{gamachiral}
\end{equation}
where $({\bf \Delta_{p}})_{ij}
\equiv \delta_{ij} - p_{i} p_{j}/p^{2}$
is the projector upon the space of transverse
polarization (normal to ${\bf p}$),
${\mathbf \Phi }$ is the
antisymmetric Hermitian tensor $\Phi _{ij}=i \epsilon _{ijk} \hat{p_{k}}$
($\epsilon _{ijk}$ being the Levi-Civita tensor). In chiral media ${\mathbf \Gamma}_{{\mathbf p}} ({\mathbf q})$ obeys the symmetry relation ${\mathbf \Gamma}_{{\mathbf p}} ({\mathbf q})
= - {\mathbf \Gamma}_{- {\mathbf p} } ({\mathbf q})
= {\mathbf \Gamma}_{-{\mathbf p}} (-{\mathbf q})$.

Our objective is to solve Eq.~(\ref{gama1}) and to explicitly
obtain the coefficients ${\mathbf \gamma}_{0}$, ${\mathbf \gamma}_{C}$ in Eq.~(\ref{gamachiral}), and a expression for natural optical activity.
To this end, let us
separate the mass operator $ {\mathbf \Sigma}_{\mathbf p} $
in a symmetric (S) and an antisymmetric (A) part:
\begin{equation}
\Delta {\mathbf \Sigma}_{\mathbf p} \equiv
- \left( \mathrm{Im} {\mathbf \Sigma}^{S}_{\mathbf p} + \mathrm{Im} {\mathbf \Sigma}^{A}_{\mathbf p} \right) \\
= - \mathrm{Im}{\mathbf \Sigma}_{\mathbf p} \left( {\mathbf U}
+ \xi {\mathbf \Phi}_{{\mathbf {\widehat p}}} \right),
\label{sigma}
\end{equation}
where ${\mathbf \Sigma}_{\mathbf p}$ is given by Eq.~(\ref{boltz1}). Equation (\ref{sigma}) defines the pseudoscalar $\xi$, whose real and imaginary parts
determines the rotatory power and the spatial dichroism of the effective medium, respectively. Inserting Eq.~(\ref{gamachiral}) into Eq.~(\ref{gama1}) and performing
the operations $ \int \frac{d {\mathbf {\widehat p}} }{ 4 \pi }
\mathrm{Tr} {\mathbf \Delta}_{{\mathbf p}}$ and
$ \int \frac{d {\mathbf {\widehat p}} }{ 4 \pi }
\mathrm{Tr} {\mathbf \Phi}_{{\mathbf p}}$
results in the two
following coupled equations for ${\mathbf \gamma}_{0}$ and ${\mathbf \gamma}_{C}$:
\begin{eqnarray}
{\mathbf \gamma}_{0} &=& 2 +
g \left(  \frac{ {\mathbf \gamma}_{0} + \xi {\mathbf \gamma}_{C}}{1+ \xi^2} \right)
+ g_{C} \left(  \frac{ {\mathbf \gamma}_{C} - \xi {\mathbf \gamma}_{0}}{1+ \xi^2} \right), \label{gamafinal1} \\
{\mathbf \gamma}_{C} &=&
- g_{C} \left(  \frac{ {\mathbf \gamma}_{0} + \xi {\mathbf \gamma}_{C}}{1+ \xi^2} \right)
- g_{CC} \left(  \frac{ {\mathbf \gamma}_{C} - \xi {\mathbf \gamma}_{0}}{1+ \xi^2} \right) \label{gamafinal2},
\end{eqnarray}
where we have put ${\mathbf q} = \hat{{\mathbf p}}$ and used the fact
that ${\mathbf  G}_{{\mathbf p}} ({\mathbf q}) \cdot {\mathbf \Delta}_{{\mathbf p}} \cdot {\mathbf  G}^{*}_{{\mathbf p}} ({\mathbf q})
= \Delta {\mathbf  G}_{{\mathbf p}} \cdot (\Delta {\mathbf  \Sigma}_{{\mathbf p}})^{-1}$. It can be shown that $g$, $g_C$, $g_{CC}$ and $\xi$ can be explicitly
written in terms of the (off-shell) scattering total T-matrix of an ensemble of $N$ point dipoles~\cite{josa}, which describes multiple scattering from the direction ${\mathbf p}$ to ${\mathbf p}^{\prime}$
\begin{equation}
{\mathbf  T}_{{\mathbf p}{\mathbf p}^{\prime}} = \sum_{N N^{\prime}} {\mathbf  S}^{N N^{\prime}} \text{exp} \left( -i \mathbf{p} \cdot {\mathbf  r}_{N} +  i \mathbf{p}^{\prime} \cdot {\mathbf  r}_{N^{\prime}}\right),
\label{tmatrix}
\end{equation}
with ${\mathbf  S}^{N N^{\prime}}$ a symmetric $3 \times 3$ matrix describing scattering of light from particle $N$ to $N^{\prime}$ that is obtained by diagonalization~\cite{dipolemodel}. As a result, multiple scattering is treated exactly, {\it i.e.} all
scattering orders are included; the only approximation within this model is at the level of single scattering, as scatterers are treated as point dipoles. Hence the knowledge of the scattering T-matrix
for the ensemble of pointlike dipoles enables one to
calculate $\xi$, $g$, $g_{C}$, $g_{CC}$ and, consequently,
${\mathbf \gamma}_{0}$ and ${\mathbf \gamma}_{C}$~\cite{josa}.
In terms of ${\mathbf  S}^{N N^{\prime}}$ and after averaging over all solid rotations,
the final expression for the pseudoscalar $\xi$, which describes
natural optical activity, is~\cite{josa}
\begin{equation}
\xi = \frac{1}{2}  \sum_{N N^{\prime}}  \text{Tr} \left\{  \right({\mathbf  S}^{N N^{\prime}} \cdot \epsilon \cdot  \hat{{\mathbf  r}}_{N N^{\prime}} \left) j_{1} \left(  k  r_{N N^{\prime}} \right)   \right\},
\label{roa}
\end{equation}
where $ j_{1} (x)$ is the spherical Bessel function of the first kind, and ${\mathbf  r}_{N N^{\prime}}$ is the relative position between scatterers $N$ and $N^{\prime}$. We have verified that Eq. (\ref{roa}) is indeed a pseudoscalar, {\it i.e.} it changes sign by a mirror operation.  We have also verified that for $N <4$, $\xi$ identically vanishes, reflecting the fact that a chiral system must be composed of at least 4 particles.

\begin{figure}[h!]
\includegraphics[width=0.4\textwidth]{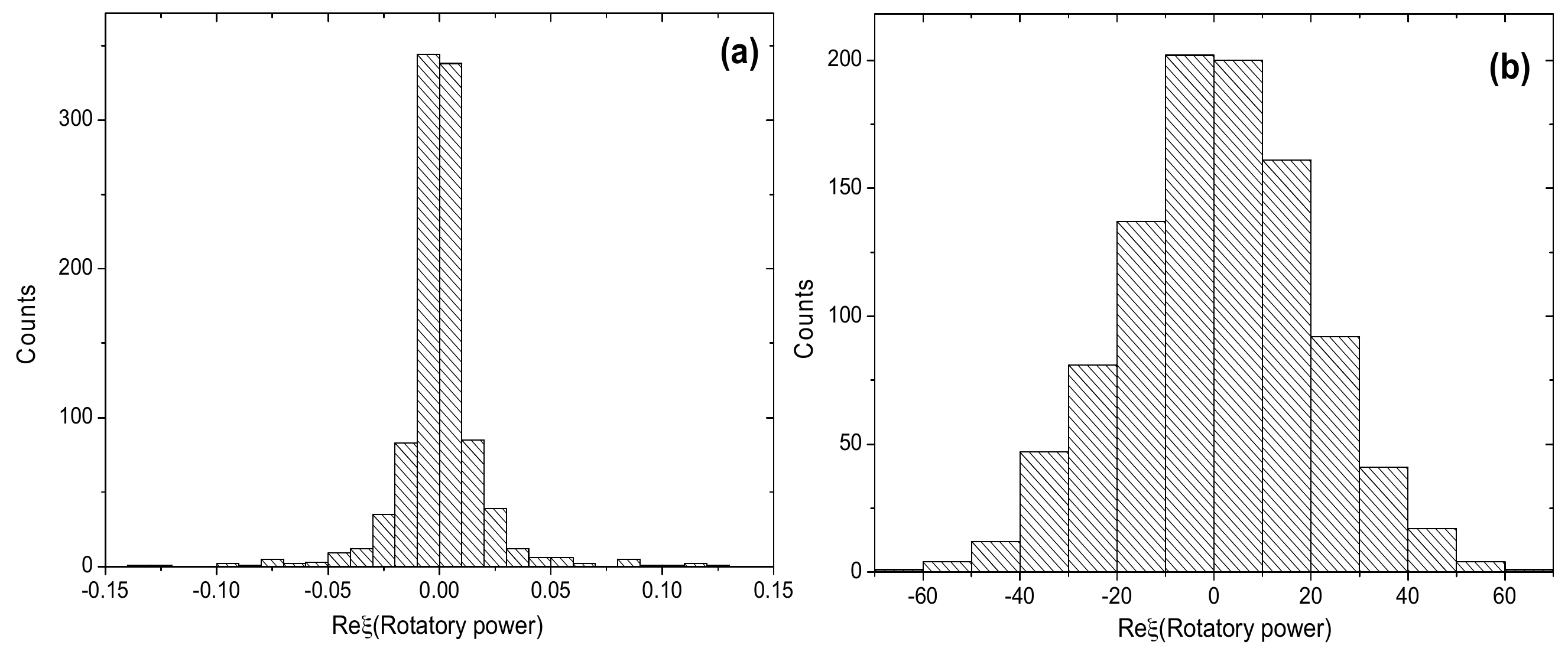} \centering
\caption{Histogram of $\textrm{Re} \xi$ that gives the rotatory power in disordered medium for (a) 10 and (b) 700 pointlike scatterers randomly distributed inside a cylinder of fixed volume (height $h / \lambda =10$ and radius $R/\lambda \simeq 5$).} \label{histogram}
\end{figure}

In Fig.~\ref{histogram} the distributions of $\textrm{Re} \xi$, which gives the optical rotatory power, is calculated using Eq. (\ref{roa}) for 1000 different disorder realizations and two different particle densities.  The scatterers are randomly distributed inside a cylinder of fixed volume (height $h / \lambda =10$ and radius $R/\lambda \simeq 5$, where $\lambda$ is the incident wavelength). In Fig.~\ref{histogram}(a) $\textrm{Re} \xi$ is calculated for $N=10$ scatterers, which corresponds to $k \ell \approx 1000$ (with $k$ the wavenumber and $\ell$ the mean free path). In Fig.~\ref{histogram}(b) $\textrm{Re} \xi$ is calculated for $N=700$ scatterers, in which case multiple light scattering is stronger ($k \ell \approx 15$). For both densities light transport in the random medium is diffusive. 
The values of $\textrm{Re} \xi$ have been normalized by the ones corresponding to $100$ scatterers distributed along a helix inside a cylinder of equal volume (see Fig.~\ref{fig1}a). Hence Fig.~\ref{histogram}b reveals that for a given configuration of random scatterers the value of rotatory power can be up to 60 times larger than for a system with the same density, where the particles are distributed along a helix, which is the hallmark of a chiral system. Figure~\ref{histogram} also shows that $\langle \xi \rangle = 0$, which can be explained by the fact that for large number of disorder realizations the mirror image of any configuration is equally probable. However, Fig.~\ref{histogram} demonstrates that the standard deviation of the distribution of $\textrm{Re} \xi$ is strongly dependent on $\ell$, which is a measure of how strong light scattering is inside the medium. We have verified that the standard deviation of $\textrm{Im} \xi$ exhibits a similar dependence on $\ell$.

\begin{figure}[h!]
\includegraphics[width=0.4\textwidth]{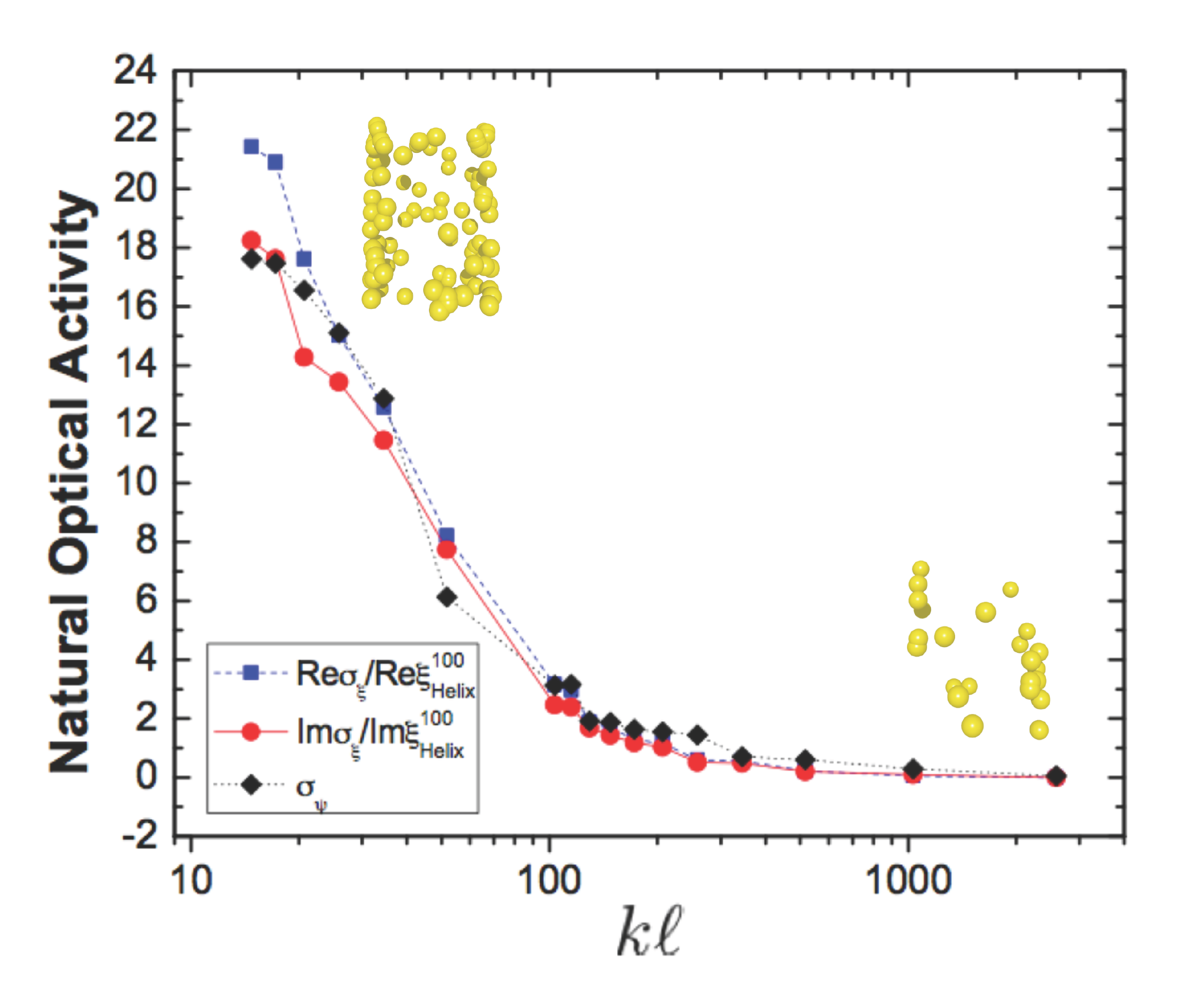} \centering
\caption{The standard deviation of the distributions of $\textrm{Re} \sigma_{\xi}$ and $\textrm{Im} \sigma_{\xi}$ for 1000 configurations of randomly distributed identical pointlike scatterers inside a inside a cylinder of fixed volume (height $h / \lambda =10$ and radius $R/\lambda \simeq 5$) as function of $k \ell$, with $k$ the wavenumber and $\ell$ the scattering mean free path. The values of $\textrm{Re} \xi$ and $\textrm{Im} \xi$ have been normalized by $\textrm{Re} \xi^{100}_{\textrm{Helix}}$ and $\textrm{Im} \xi^{100}_{\textrm{Helix}}$, which correspond to rotatory power and spatial dichroism of 100 pointlike dipoles distributed along a helix inside a cylinder of the same volume. Also shown is the standard deviation $\sigma_{\psi}$ of the distributions of chiral geometrical parameter $\psi$, defined in Ref.~\cite{harris}, of the corresponding 1000 random configurations. The insets show a scheme of characteristic random configurations for high particle density (small $k \ell$) and for low particle density (large $k \ell$).} \label{roakl}
\end{figure}

Figure~\ref{roakl} shows the real ($\textrm{Re} \sigma_{\xi}$) and imaginary ($\textrm{Im} \sigma_{\xi}$) parts of the standard deviation of $\xi$ for 1000 disorder realizations as a function of $k \ell$ for a diffusive scattering system ($10  \lesssim k \ell  \lesssim 3000$) composed of resonant pointlike scatterers randomly
distributed inside a cylinder of fixed volume (height $h / \lambda =10$ and radius $R/\lambda \simeq 5$). The results are normalized by the value of $\xi$ for a system composed of 100 scatterers distributed along a helix, a system that is clearly chiral, contained in a cylinder of equal volume. This allows us to obtain the order-of-magnitude of natural optical activity of a disordered medium, as natural optical activity for nanoparticles oriented along a helix has been calculated in Ref.~\cite{zan}.
Figure~\ref{roakl} reveals that $\sigma_{\xi}$  monotonically increases as $k \ell$ decreases, showing that the magnitude of natural optical activity increases as the density of scatterers increases. More importantly, for $k \ell \lesssim 20$ $\sigma_{\xi}$ can be approximately 15 times larger than the case of a helix. It is also important to emphasize that both  $\textrm{Re} \sigma_{\xi}$ and  $\textrm{Im} \sigma_{\xi}$ are small for ballistic and weakly scattering systems, for which $ k \ell \gg 1$. This demonstrates that multiple light scattering is necessary in order to produce significant values of natural optical activity. In particular, we have verified that at least four scattering events are required to generate a non-vanishing value of $\xi$, so that light can ``probe'' the chiral configuration associated to random scatterers. Furthermore, in the diffusive regime, in which the system size is much larger than $\ell$, the statistical distribution of $\xi$ is not expected to depend on the system size for a large set of disorder realizations.  

To further investigate the interplay between the chirality associated with the random configuration of scatterers and natural optical activity, we calculate a geometrical chiral index $\psi$ defined in Ref.~\cite{harris} according to
\begin{equation}
\label{psi}
\psi = Q^{il} B^{jm} \epsilon_{ijk} S^{klm},
\end{equation}
with $\epsilon_{ijk}$ the Levi-Civita tensor. In Eq.~(\ref{psi}) the tensors $S$, $B$, and $Q$ are given by
\begin{eqnarray}
S^{klm} &=& \sum_{\alpha} \left[ r^{k}_{\alpha} r^{k}_{\alpha} r^{l}_{\alpha} - \frac{1}{5} \left(r_{\alpha}\right)^{2} \left(r^{k}_{\alpha} \delta^{lm} + r^{l}_{\alpha} \delta^{km}  + r^{m}_{\alpha} \delta^{kl} \right) \right], \nonumber \\
B^{ij} &=& \psi_{B} \left( e^{i}_{1}e^{j}_{1} - e^{i}_{2}e^{j}_{2} \right) \equiv \psi_{B} \widetilde{B}^{ij}, \nonumber \\
Q^{ij} &=& \psi_{Q} \left( e^{i}_{3}e^{j}_{3} - \frac{1}{3} \delta^{ij} \right) \equiv \psi_{Q} \widetilde{Q}^{ij},
\label{tensors}
\end{eqnarray}
where $\textbf{r}_{\alpha}$ is the position vector of the scatterers $\alpha$ relative to the system center of mass, and $\psi_{B}$ and $\psi_{Q}$ are the eigenvalues corresponding to the matrices $\widetilde{B}$ and $\widetilde{Q}$, respectively.
The chiral index $\psi$, which only depends on the scatterers positions, is a pseudoscalar invariant under rotations that vanishes for achiral configurations~\cite{harris}. The choice of chiral index is not unique and other chiral measures do exist~\cite{chiralindexes}. The chiral index~(\ref{psi}) has been chosen here by its simplicity and, more importantly, because it has been shown to be related to physical observables, such as the magneto-chiral scattering cross-section~\cite{pinheiro2002}, the electronic current in chiral carbon nanotubes~\cite{pinheiro2010}, and the pitch of a cholesteric liquid crystal~\cite{harris}.
For a random ensemble of scatterers the average value of $\psi$ over many disorder realizations is zero since the mirror image of any configuration is equally probable. However, the standard deviation $\sigma_{\psi}$ is strongly dependent on the particle density. This can be seen in Fig.~\ref{roakl}, where $\sigma_{\psi}$ is calculated using Eq.~(\ref{psi}) for 1000 disorder realizations. Remarkably, Fig.~\ref{roakl} shows that $\sigma_{\psi}$  is proportional to $\sigma_{\xi}$, confirming that the calculated optical activity is indeed related to the intrinsic chiral nature of disordered systems. Together with previous experimental evidence of optical activity in disordered systems in the absence of any chiral substance~\cite{hadley,silverman,drachev}, these results strongly suggest that large optical activity should be observed in a disordered medium due to its intrinsic chirality. As a possible application, Fig.~\ref{roakl} suggests that one could determine the particle density in solution by measuring natural optical activity.

\begin{figure}[h!]
\includegraphics[width=0.4\textwidth]{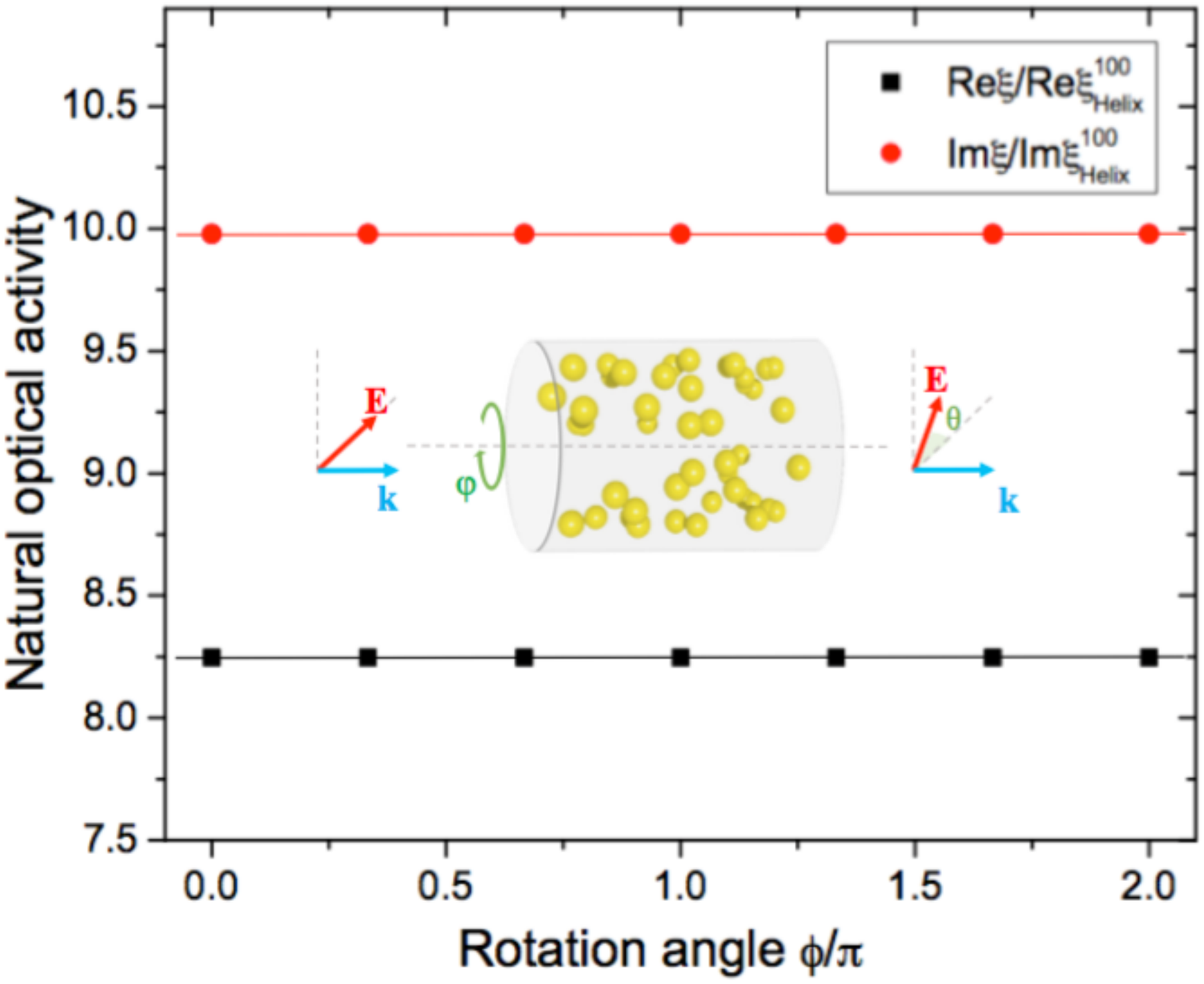} \centering
\caption{$\textrm{Re} \xi$ and $\textrm{Im} \xi$ for a fixed random configuration of 100 dipoles inside a cylinder of same volume as a function of the rotation angle $\phi$ around the cylinder main axis. The values of $\textrm{Re} \xi$ and $\textrm{Im} \xi$ have been normalized by  $\textrm{Re} \xi^{100}_{Helix}$ and $\textrm{Im} \xi^{100}_{Helix}$ that correspond to the corresponding values of $\xi$ for 100 scatterers distributed along a helix inside a cylinder of the same volume.} \label{anisotropy}
\end{figure}

In order to investigate whether the calculated natural optical activity is related to the anisotropy of the disordered medium, we have calculated the real and imaginary parts of $\xi$ for a single, fixed configuration of random scatterers inside a cylinder with the same dimensions of Fig.~\ref{histogram}, which is continuously rotated by an angle $\phi$ around the main axis of the cylinder. The result, shown in Fig.~\ref{anisotropy}, reveals that $\xi$ does not depend on $\phi$. Hence we rule out the possibility that the reported results are due to the anisotropy of the random medium as Fig.~\ref{anisotropy} proves. Indeed, polarization change due to anisotropy does not require multiple light scattering. On the contrary, here optical activity is related to the intrinsic chirality of random media, which light can only probe when multiple scattering occurs (or, more precisely, when at least four scattering events occurs), as Fig.~\ref{roakl} demonstrates.

\begin{figure}[h!]
\includegraphics[width=0.4\textwidth]{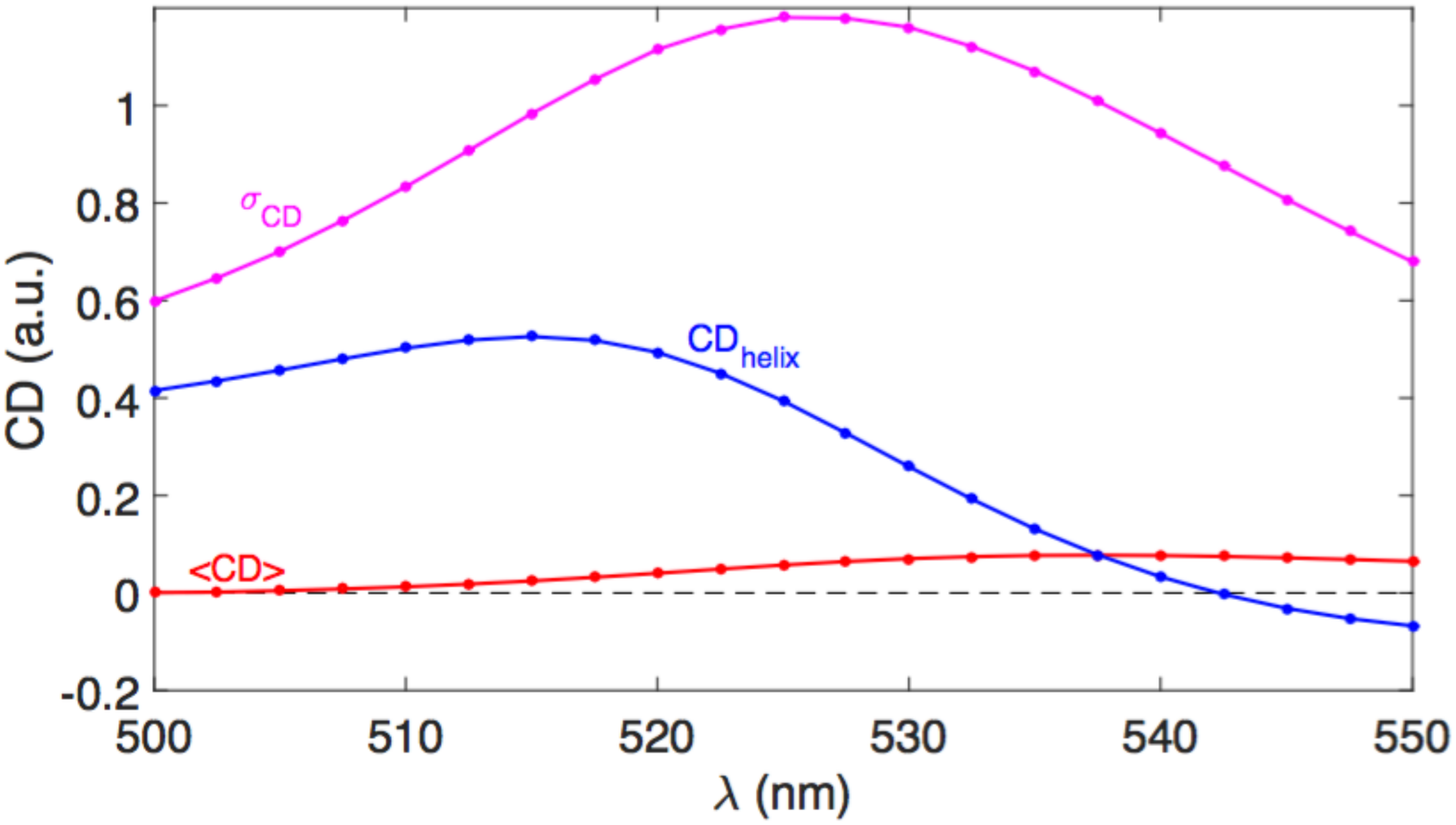} \centering
\caption{Circular dichroism (CD) spectra of random and helical Au nanoparticle configurations. Blue line represents the CD of a helical arrangement with radius$=467$ nm \& pitch$=660$ nm. The average, $\langle CD \rangle$, over 310 realizations of random nanoparticle configurations is represented by the red line. Magenta line represents the standard deviation $\sigma_{CD}$ of CD. The radius of the nanoparticles is considered to be 25 nm. All CD spectra have been obtained by finite element modeling.}\label{fullwave}
\end{figure}

To further confirm the validity of the {results from the microscopic theory of light transport, we performed finite element full-wave simulations of random and helical configurations of achiral Au nanoparticles. In Fig. 5, we show circular dichroism (CD) as a function of wavelength for random (red and magenta lines) and helical (blue lines) ensembles of 25 nanoparticles. Both random and helical configurations occupy the same volume at a density of $\simeq 4$ nanoparticles/$\lambda^3$. The helical configuration exhibits strong circular dichroism in the vicinity of the plasmonic nanoparticle resonance ($\simeq 525$ nm). On the other hand,  in the case of random nanoparticle ensembles, the average over 310 realizations is close to zero, as predicted by the previous microscopic analysis. However, the corresponding standard deviation ($\sigma_{CD}$) exceeds the CD of the helical configuration by a factor of 2 at resonance, while for specific realisations this factor can be as high as 10 (not shown here). These results are in qualitative agreement with the predictions of the point-dipole model for diffusive scattering systems, and demonstrate unambiguously that polarization effects in disordered scattering systems can be much stronger than in a typical chiral medium (helix) due to its intrinsic chirality.

In conclusion, we have investigated natural optical activity of diffusive disordered systems composed of non-chiral plasmonic resonators. Using microscopic electromagnetic wave transport theory coupled to vectorial Green's matrix model for pointlike scatterers, we have demonstrated that the distributions of the rotatory power and spatial dichroism in diffusive disordered media
is strongly dependent on the scattering mean free path. In particular, despite the fact the mean value of rotatory power and spatial dichroism in random media vanishes after averaging over many disorder realizations, their standard deviation exhibits a strong monotonic increase as the scattering mean free path decreases. As a result we argue that the standard deviation of both rotatory power and spatial dichroism is the appropriate quantity to probe chirality and natural optical activity in disordered systems of non-chiral plasmonic oscillators. This finding is independently corroborated by full-wave electromagnetic calculations. We have found that the standard deviation of a purely geometrical chiral parameter, defined in Ref.~\cite{harris} and that only depends on the scatterers' positions, is proportional to the standard deviation of natural optical activity in random media. This result corroborates that the latter is the appropriate quantity
to probe natural optical in disordered media, which is intrinsically related to the fact that random systems lack mirror symmetry.

We thank B. A. van Tiggelen, L. Dal Negro, and O. Buchnev for fruitful discussions. We thank The Royal Society-Newton Advanced Fellowship (Grant no. NA150208). F.A.P. thanks the Optoelectronics Research Centre and Centre for Photonic Metamaterials, University of Southampton, for the hospitality, and CAPES for funding his visit (Grant No. BEX 1497/14-6). F.A.P. also acknowledges CNPq (Grant No. 303286/2013-0) for financial support. N. P. acknowledges the support of the UK's Engineering and Physical Sciences Research Council (Grant no. EP/M008797/1).


\end{document}